\documentclass[twocolumn]{article}
\usepackage{graphicx} 
\usepackage{hyperref}
\usepackage{abstract}
\usepackage{amsmath, amssymb}
\usepackage{geometry}
\usepackage{booktabs}
\usepackage{amsfonts}
\usepackage{booktabs}  
\usepackage{multicol}    
\usepackage{caption}   
\usepackage{titling}
\title{On the hidden costs of passive investing}
\author{Iro Tasitsiomi}
\date{June 2025}

\begin{document}
\twocolumn[
\maketitle

\begin{center}
\begin{minipage}{\textwidth}
\begin{abstract}
 Passive investing has gained immense popularity due to its low fees and the perceived simplicity of focusing on zero tracking error, rather than security selection. We consider the system of a passive asset manager and a trader in the context of index reconstitution. Focusing on zero tracking error, passive asset managers often execute index reconstitution trades at the market close of the day the index vendor makes the change, whereas liquidity providers like the trader accumulate the stock ahead of this rebalancing. We consider various inventory paths, as well as solve for the optimal such paths under both Nash (simultaneous decisions) and Stackelberg (trader leads, manager responds) dynamics. Our analysis shows that the passive (zero tracking error) approach results in costs amounting to hundreds of basis points compared to strategies that involve gradually acquiring a small portion of the required shares in advance, with minimal additional tracking errors. In addition, we show that under all scenarios analyzed, a trader who builds a small inventory post-announcement and provides liquidity at the reconstitution event can consistently earn several hundreds of basis points in profit, and often much more, assuming minimal risk.
\end{abstract}


\end{minipage}
\end{center}]

\section{Introduction} 
Market indices such as the S\&P 500, FTSE 100, and MSCI benchmarks are dynamic: to maintain accurate representation of their target market segments, securities are periodically added and removed.  A large literature documents abnormal return patterns around such index reconstitution events (e.g., \cite{chen2004, lynch2001, madhavan2000, chang2015, sammon2025, hendrix2024, li, arnott, chinco, reilly, micheli}). 

More specifically, \cite{chang2015} show that Russell index inclusions trigger immediate and sizable price effects not explained by fundamentals. \cite{sammon2025} and \cite{hendrix2024} find that mechanical rebalancing exposes index funds to adverse selection, slippage, and buy-high/sell-low cycles. \cite{li} estimates that the execution cost at the close of the index change day is higher by $\sim 67 \rm{bps}$ (trade-size weighted average cumulative adjusted return (CAR) on stocks being traded by those ETFs, adjusted for Fama-French factors) making it three times larger than the average cost incurred for similar trades under no reconstitution circumstances. Similarly, \cite{arnott} find that from the announcement date to the market close on the trade date, index additions on average appreciate by $\sim 5.0\%$ relative to the market, while deletions trail the market by $\sim 7.2\%$. Furthermore they find that these price movements are highly statistically significant.  

Evidence suggests that passive flows exacerbate these effects: \cite{israeli2017} show ETF activity distorts shorting demand and liquidity; \cite{bendavid2018,  appel2016} document that greater ETF ownership raises volatility and non-fundamental trading pressure; and \cite{crane2016} demonstrate that passive ownership via index thresholds can mechanically influence payout policy beyond pure price impact.

One of the reasons for such abnormal patterns is that large public indices announce reconstitutions several days prior to the implementation date, while an estimated $\sim 56\%$ \cite{li} of passive ETFs place much of the trading positions at the closing price of the reconstitution day. These passive investors wait until the exact moment that the index change occurs so that they do not incur tracking error with respect to the benchmark. But, the rest of the market can participate in anticipatory trading of the name that has been announced.  Such anticipatory trading from market players can occur days, weeks, or even months before the index rebalancing event (i.e., even prior to announcement if the trader thinks they have a good idea of what the index changes will be), but it is more prevalent and very low risk after announcement when the name to be traded is known. 

Of course, other factors, such as improved analyst coverage or an increase in expected future liquidity, may contribute to these price moves, in the case of additions. Another important feature that potentially contributes to the return dynamics is the lower risk profile of the larger and often more familiar companies added to indices and the higher risk profile of the deletions. However, all these should be captured in the post reconstitution price, but as \cite{li} and  \cite{arnott} show, most of the price change between announcement and reconstitution is actually reversed post-reconstitution. 

Thus the question that arises is: what is the cost of adhering strictly to zero tracking error? Specifically, when asset managers wait until the close of the reconstitution day to fully adjust, what premium do they pay compared to strategies that involve modest early trading with minimal tracking error?

We should note that there are various ways to look at index reconstitutions. E.g., \cite{madhavan22} propose that price changes should be interpreted as arising from net flows aggregated across all indices. This framework explicitly accounts for index migrations: a stock may exit one Russell index while simultaneously entering another, in which case the overall flow impact depends on the stock’s market capitalization and on the assets under management (AUM) tracking each of the two indices. The authors also aggregate flows across managers - and since buys and sells will have to (roughly at least) offset at the market level, they find that the net liquidity provision (albeit defined a bit different than here) is "small overall". 

However, for most investors in passive funds, this aggregation on multiple levels is mostly theoretical. They typically own only a segment of the market, often through specific asset managers or mandates, and therefore will not benefit from offsetting flows with other managers or indices... And, even for large passive managers, internal crossings are far from perfect and come with several regulatory and operational constraints. So even though this omniscient observer approach makes sense on a market level, it is not the view all investors and mandates have available to them.
 
This paper aims at providing frameworks for plausible modelling and estimates of this premium under different assumptions/conditions. We model the interaction between a passive index-tracking manager and a strategic trader using game-theoretic frameworks. We focus on the cost implications of execution timing under Nash and Stackelberg dynamics and we quantify, in a stylized but tractable model, the hidden implementation costs that arise when a manager prioritizes zero tracking error at the expense of optimal execution. We model price dynamics following prior seminal works (\cite{almgren2000optimal}), we contribute to the growing literature on the microstructure implications of passive investing and provide relevant insights for optimal execution.  

We assume that there is effectively only one asset manager (who needs to buy the full reconstitution volume). Any security demand not met by our one trader, is assumed to be met at reconstitution by other passive market participants who happen to hold some of the specific security (say multiple such investors that had bought some of this security at some point, irrespective of the reconstitution event). Future extensions of this work will include more competing traders with even richer dynamics (along the lines of, e.g., \cite{chriss2024competitiveequilibriatrading, chriss2025positionbuildingcompetitiongame}), as well as allow for post-reconstitution trading. 

Last but not least the empirical observations and results that have inspired this work are taken as given and have been cited in this introduction. It is not in scope for this paper to make a case for them. Furthermore, the market is a very complex system, and the claim is not that our relatively simple frameworks will capture all scenarios happening. To show the usefulness of the frameworks discussed in providing realistic approaches and estimates, we provide a representative predicted price path in the discussion. 

\section{The formulation}
\subsection{Price Dynamics}
We denote by:
\begin{itemize}
    \item  $S(t)$: the price of the stock that was announced at time 0 and will be added to the index at the close of the market N days from the announcement, at time $t$, for $t \in [0, t_N]$.  
    \item $v(t)$: net trading rate (shares per unit time)
    \item $\gamma$: permanent impact coefficient (change in price per share per unit time). In the case of index reconstitution, permanent means persisting (at least) until the time of reconstitution.
    \item $W(t)$: Brownian standard motion
    \item $\sigma$: volatility
    \item $\mu$: drift of the fundamental price
\end{itemize}

The price process is then given by:
$$
dS (t) = \mu, dt + \sigma, dW (t) + \gamma v (t), dt
$$

Both here, and in what follows, we borrow from \cite{almgren2000optimal} - the reader is referred to that work for a discussion around the simplification of using an arithmetic Brownian motion for $S(t)$, and other modeling assumptions. 
\subsection{Execution Price}
The actual execution price $\tilde{S}(t)$ also includes the temporary impact:
$$
\tilde{S}(t) = S(t) + h(v(t)) = S(t)  + \eta A(t) \pm \epsilon $$
where $h(v) = \eta A(t) \pm \epsilon$ is the temporary (transient) impact with $\epsilon$ the fixed trading cost (say, half of the bid-ask spread), $A(t)$ is the cumulative residual temporary price impact at time $t$ measured in number of shares traded per unit time and $\eta$, measured in $USD \cdot sec / shares^2$, is the temporary impact coefficient that scales the impact to USD per share. For simplicity, we will use $A(t)=v(t)$, namely that temporary price impacts only affect the current time step - there are no residual such impacts from prior times/trading.

We now derive the actual execution price at time $t$ from the initial price $S_0 = S(0)$. For the fundamental price evolution (including permanent impact), integrating from $0$ to $t$:
$$
S(t) = S_0 + \mu t + \sigma W(t) + \gamma \int_0^t v(s)\,ds
$$
This is the price that the market observes (excluding the temporary impact), influenced by the net cumulative flow up to time $t$.

Substituting $S(t)$ from above into the execution price formula, we get:
$$
\tilde{S}(t) = S_0 + \mu t + \sigma W(t) + \gamma \int_0^t v(s)\,ds  + \eta v(t) \pm \epsilon 
$$
which is the final, most general expression of the execution price at time t. 

In what follows for simplicity we will omit the drift term - the amount of time involved are small, anyway (between stock index inclusion announcement and reconstitution) - as well as the randomness term that would anyway sum to 0 on expectation, so our final execution price expression becomes:
\begin{equation}
\tilde{S}(t) = S_0 + \gamma \int_0^t v(s)\,ds + \eta v(t) \pm \epsilon
\end{equation}
\subsection{Possible scenarios, assumptions and benchmark definitions}
In what follows we denote by $x(t)$ the aggregate number of shares the trader has bought by time $t$, starting at $t=0$, right after index inclusion of the specific stock has been announced. We assume the trader accumulates a total of $T = \int_0^N \dot{x}(t)\,dt = x(t_N)$ shares, a number that is constrained by the capital the trader has available, the price evolution, and, possibly, by the desired target return levels and by her ability to forecast $D$, i.e., the total liquidity the asset manager will need to implement the index change. It should be noted that $T$ is the number of shares the trader appears with at reconstitution ready to sell to the asset manager, but we do not require that $x(t) \leq T$. Indeed, we will see some market manipulation behaviors where the trader buys at some point many more shares than $T$ to push the price higher up and maximize her profit. 

We denote by $y(t)$ the aggregate number of shares bought by the asset manager by time $t$. Denoting by $D'$ the shares the manager bought from other market participants (in small chunks/rates) and using the equation for the execution price, we have:
\begin{equation*}
\tilde{S}(t_N-) = S_0+ \gamma(T + D') + \eta(\dot{x} (t_N) + \dot{y} (t_N)) \pm \epsilon
\end{equation*}
where the $-$ at $t_{N}$ it to indicate the "just before exactly $t_N$".
At exactly $t_N$ the permanent price impact term goes from $T+D'$ to $T+D'-T+D-D'=D$ and the temporary term $[-T + (D-D')] \cdot \delta(t-t_N)$ :
\begin{equation*}
\tilde{S}(t_N) = S_0+ \gamma D + \eta [-T + (D-D')] \cdot \delta(t-t_N)\pm \epsilon
\end{equation*}
with the $\delta$ function reminding us that often managers try to buy what they need at the close. Thus, we set our benchmark as the case where $D'=0$. This "buy all at once at reconstitution" price is what we will be referring to as the "benchmark" price from this point on:
\begin{equation*}
\tilde{S}(t_N) = S_0+ \gamma D + \eta (D-T) \cdot \delta(t-t_N) + \epsilon
\end{equation*}
for a total benchmark cost of:
\begin{equation}
Cost_{bench} = D [S_0+ \gamma D + \eta (D-T)\cdot \delta(t-t_N) + \epsilon]
\end{equation}

In what follows, we put our numerical results in perspective by comparing with this benchmark: asset manager "savings" are calculated as this benchmark acquisition cost minus that of the whatever strategy the asset manager follows. It should be noted that the magnitude of this benchmark cost depends on how one decides to treat the "instantaneous" derivative at $t_N$ - it can be very sensitive to the temporary impact term: how fast one buys matters a lot- buying $D-T$ shares throughout a day can yield very different prices than buying (/trying to buy) them all at the close of the $t_{N}$th day, exactly when the index changes. From a mathematical point of view, it comes down to the $\delta t$ choice in which the $D-T$ shares are bought since it determines the rate $(D-T)\cdot \delta(t-t_N) \sim (D-T)/ \delta t$.

We assume $\delta t$ = 1 day, that is, our benchmark cost is the cost that our model predicts if the asset manager bought the $D-T$ shares throughout day $t_N$ at a constant rate. That means that this is a conservative (harder to beat) benchmark compared to "buying all at market close" - a smaller $\delta t$ would lead to really big temporary price impacts. That is why in the results we share later, as the asset manager weighs more and more the tracking error rather than the cost, we start seeing negative asset manager "savings": the trade starts resembling more and more a spike purchase towards the close of the reconstitution day.

\section{Objectives}
\subsection{The Trader}
Denote by $x(t)$ the aggregate number of shares the trader has bought by time $t$, starting at $t=0$, right after index inclusion of the specific stock has been announced. Also assume the trader accumulates a total of $T = \int_0^N \dot{x}(t)\,dt = x(t_N)$ shares, a number  that is constrained by the capital the trader has available, as well as by her ability to forecast $D$. Lastly, the trader sells all her $T$ shares at the market price at the close on day $N$. The total cost of buying these shares is:

$$
 \int_0^{t_{N}} \left[ S_0+ \gamma(x + y) + \eta(\dot{x} + \dot{y}) + \varepsilon \right] \dot{x} (t) dt$$
with $y(t)$ the number of shares that the asset manager has bought over time $t$ for $y(t_N) = D$.

At $t_N$ the trader sells all her $T$ shares to the asset manager at price $\tilde{S}(t_N)$ and receives a total of:
$$
T \cdot \tilde{S}(t_{N^+}) = T \cdot (S_0+\gamma D - \epsilon )
$$
The $+$ at $t_N$ is to indicate that we assume $\tilde{S}(t_N)$ without the temporary impact term. We assume that because of both the relative arbitrariness of this term (albeit we could make the same assumption as for the manager), and because we do not want to "reward" the trader for any mismatch of the asset manager demand $D$. Or, alternatively, one can look at this choice as choosing a conservative approach in calculating trader profits.

So the trader's objective is to minimize the functional:

\begin{align}
J_T[x] &  = \int_0^{t_{N}} \left[S_0+ \gamma(x + y) + \eta(\dot{x} + \dot{y}) + \varepsilon \right] \dot{x}(t) \, dt \nonumber \\
& - T \cdot (S_0+ \gamma D - \epsilon)   
\label{eq3}
\end{align}

\subsection{The Asset Manager}
Our trader will be buying $T$ shares over $[0, t_{N}]$ to front-run anticipated index demand $D$. The index manager needs to buy $D$ shares and seeks to track the index closely, but may incur high costs by buying all at the close (as prices have been permanently impacted). Thus in general, the asset manager would try to balance:
\begin{enumerate}
    \item The tracking error: deviation from market cap-weighted benchmark (if the stock is added at close but manager buys gradually).
    \item The execution cost: trading earlier may reduce cost but increases tracking error.
\end{enumerate}

For what follows, let's define as
\begin{itemize}
    \item $A$: Total AUM of the fund
    \item $y(t)$: Number of shares of the new security held by the fund at time $t$
    \item $S(t)$: Price per share of the new security at time $t$
\end{itemize}

For simplicity we make the following assumptions:
\begin{itemize}
   \item The asset manager purchases new shares with funds outside the index tracking fund. 
   \item The fund AUM is significantly larger than the new money/stock addition so that we can assume $A$ constant. 
   \item There is only one name addition - no deletions, etc.
\end{itemize}
The weight of the new security in the fund will be:
  $$
  w_{\text{a}}(t) = \frac{y(t) \cdot S(t)}{A}
  $$
This weight in the index, $w_{bench}$ will be

  $$
  w_{\text{bench}}(t) = 
  \begin{cases}
    0, & t < t_N \\
    \frac{D \cdot S(t_N)}{A}, & t = t_N
  \end{cases}
  $$

\subsubsection{Tracking Error}
The core of modeling tracking error in index reconstitution is the timing mismatch between when the manager trades and when the index actually changes. We want to penalize the manager for starting to accumulate $y(t) > 0$ before $t_N$, instead of buying everything at the close of trading $t_N$. This reflects the tracking error versus a benchmark that holds zero until $t = t_N$ and then jumps to $D$ instantly at the close, which is often the actual approach followed. 

We denote by $\lambda$ the penalty/risk aversion coefficient (units of 1/time) and 

We denote by $V$ the covariance matrix of the returns. The tracking error will be as follows:

\begin{align*}
\Delta TE^{2} (t) = (w_{\text{a}}(t) - w_{\text{bench}}(t))'V (w_{\text{a}}(t) - w_{\text{bench}}(t))
\end{align*}

The first term in parentheses is the difference between vectors $(w_1,w_2,...,w_{a})-(w_1,w_2,...,w_{bench})$ which has only one nonzero element. The same is the case with the column vector in the last parenthesis. The 0s cancel any V terms that would involve covariance with other returns. As such, the expression becomes: 
\begin{align*}
\Delta TE^{2}(t) = (w_{\text{a}}(t) - w_{\text{bench}}(t))^{2} \sigma^2  
\end{align*}
where $\sigma^2$ is the return variance of the new stock. 
Substituting the expressions for the weights we get, for $t < t_N$:
\begin{equation}
\Delta TE^2 (t) = \left( \frac{y(t) \cdot S(t)}{A} \right)^2 \cdot \sigma^2
\end{equation}

For $t \geq t_N$ we assume that $\Delta TE$ is 0 since we are focusing on tracking error because of pre-reconstitution trading. 

Multiplying and dividing by $D \cdot S(t_N)$, we obtain:
$$
\Delta TE^{2}(t) = \left (\frac{y(t)}{D}\right)^{2} \cdot w_{bench}(t_N)^2 \sigma^2
$$

For simplicity, and for higher tractability we assumed $S(t) \simeq S(t_N)$. The expression is expected to be an upper limit to $\Delta TE$ since we expect $S(t) \leq S(t_N)$. This assumption keeps our TEP quadratic in the number of shares.

In the appendix we show how starting more from an ex post, rather ex ante TE mindset - by roughly quantifying the active return of the fund compared to the index - we arrive at the same expression.

For the total $\Delta TE^2[y]$ over the $t_N$ days from index inclusion announcement to reconstitution, we have the following:

\begin{equation}
\Delta TE^2[y] = w_{bench}^2(t_N)\cdot \sigma^{2} I[y]
\end{equation}
with
\begin{equation}
I[y]=\int_0^{t_N} \left( \frac{y(t)}{D} \right)^2 dt
\end{equation}

This expression for the tracking error has many desirable properties:
\begin{itemize}
    \item Is always non-negative,
    \item Is zero only if $y(t) = 0$ for all $t < t_N$,
    \item Penalizes the manager for accumulating inventory early.
\end{itemize}
In addition, this expressions is smooth and differentiable and as we will see leads to analytic and numerically tractable solutions. 

It should be noted that a. this is already annualized and b. the volatility $\sigma_r$ is daily volatility (so it is equal to the annual divided by $\sqrt{252}$). 
\subsubsection{Cost and final objective}
The cost for the asset manager is given by a similar expression as the trader cost by replacing $\dot{x}$ with $\dot{y}$ outside the square brackets. Penalizing for tracking error makes the functional that the manager should try to minimize:
\begin{align}
J_A[y] & =\int_0^{t_N} \left[ S_0+ \gamma(x + y) + \eta(\dot{x} + \dot{y}) + \varepsilon \right] \dot{y}(t) \, dt \nonumber \\
&  + \lambda \cdot I[y]
\end{align}
where $\lambda$ the parameter that determines the balance between cost and tracking error. It should be noted that the 2 terms have significantly different value ranges - orders of magnitude different. To ensure the relative competition of the 2 effects - cost and tracking error - assume that the second is multiplied by the "vanilla" cost, $Cost_{bench}$ (or that the first term is divided by it). This is not the only way to force something like this, but it is probably the simplest, and it is independent of any model that we consider in this work. 

In what follows, when we are calculating savings for the asset manager we mean the quantity
\begin{align*}
Savings & = Cost_{bench} - \\
 & \int_0^{t_N} \left[ S_0+ \gamma(x + y) + \eta(\dot{x} + \dot{y}) + \varepsilon \right] \dot{y}(t) \, dt 
\end{align*}
and we convert to basis points by dividing the savings by $Cost_{bench}$ and multiplying by 10000.

Lastly, we left $w_{bench}$ and $\sigma_{r}$ outside of this objective. In some sense they can be looked at as absorbed by $\lambda$ and after optimizing once, we can see what the various $\Delta TE$ can be for different assumptions for stock volatility and index inclusion weight.
 \begin{table*}[t]
\centering
\begin{tabular}{|l|c|c|c|c|}
\hline
\textbf{Parameter} & \textbf{Large-Cap} & \textbf{Mid-Cap} & \textbf{Small-Cap}  & \textbf{Values Used}\\
\hline
Stock Price (\$) & 100 & 50 & 30 & 50 \\
Reconstitution Volume & 10M – 20M & 2M – 5M & 500K – 2M  & 5M \\
Index Weight & 1\% & 0.2\% & 0.05\% & 1\% \\
Volatility \( \sigma \) (annualized) & 25\% & 35\% & 50\%  & 30\% \\
Permanent Impact \( \gamma \) (USD/share\(^2\)) & \(1 \times 10^{-7}\) & \(5 \times 10^{-7}\) & \(1 \times 10^{-6}\) & \(10^{-7}\) \\
Temporary Impact \( \eta \) (USD·s/share\(^2\)) & \(1 \times 10^{-6}\) & \(5 \times 10^{-6}\) & \(1 \times 10^{-5}\)  & \( 10^{-6}\) \\
Half-Spread \( \varepsilon \) (USD/share) & 0.005 & 0.02 & 0.05 & 0.01 \\
\hline
\end{tabular}
\caption{Real world model parameters by market capitalization and values the results we share mostly focus on.}
\label{tab:params}
\end{table*}

\section{Euler–Lagrange Equations}

We treat $x(t), y(t)$ as control variables and use:

$$
\frac{d}{dt} \left( \frac{\partial \mathcal{L}}{\partial \dot{z}} \right) - \frac{\partial \mathcal{L}}{\partial z} = 0
\quad \text{for   }  z = x, y
$$
 with $L$ the Lagrangian.
\subsection{Trader}
The Lagrangian is
$$
\mathcal{L}_T = \left[ S_0 +\gamma(x + y) + \eta(\dot{x} + \dot{y}) + \varepsilon \right] \dot{x}
$$
where we have already omitted the constant term of revenue received when selling all $T$ shares at reconstitution. For
$$
\frac{\partial \mathcal{L}_T}{\partial \dot{x}} = S_0 + \gamma (x + y) +\eta \dot{y} + \epsilon + 2\eta \dot{x}
$$

and then differentiating with respect to time:

$$
\gamma \dot{x} + \gamma \dot{y} + \eta \ddot{y} + 2\eta \ddot{x}
$$
minus

$$\frac{\partial \mathcal{L}_T}{\partial x} = \gamma \dot{x}
$$

gives us:
\begin{equation}
2\eta \ddot{x}(t) + \eta \ddot{y}(t)+ \gamma \dot{y}(t) = 0
\end{equation}

It is worth mentioning that for $y(t)=0$, we simply recover the solution for optimal trading (although their solution is for liquidation) by \cite{almgren2000optimal}.

\subsection{Asset Manager}
In this case, the Lagrangian is

$$
\mathcal{L}_A = \left[S_0+ \gamma(x + y) + \eta(\dot{x} + \dot{y}) + \varepsilon \right] \dot{y} + \frac{\lambda}{D^2} y^2
$$

Following the same procedure as before, one can show that the resulting equation is:
\begin{equation}
\gamma \dot{x} + \eta \ddot{x} + 2\eta \ddot{y} - 2 \frac{\lambda}{D^2}y =0
\end{equation}

\section{Model Parameters}

For this work, we want to motivate high-level estimates for the plausible range of the various coefficients and parameters, not to calculate them accurately. Choices we find reasonable are shown in Table ~\ref{tab:params}.

The values of our price impact parameters were motivated from the empirical studies that have also motivated our price evolution model(e.g., \cite{almgren2000optimal, gatheral2010, bacry2015, bershova2013}). Values these studies would suggest, depending on the market cap, are shown in Table ~\ref{tab:params}. For $\gamma$, we choose 1 order of magnitude less than $\eta$. 

Our reconstitution volume and index weights were motivated by multiple sources - see, e.g., \cite{madhavan2003, chen2006, beneish2002, petajisto2011, ftserussell, nasdaqquality, taqdata}. 

We assume there are 10 trading days between index inclusion announcement and index reconstitution. This is somewhat in the middle of the S \& P indices where this is closer to a few business days and the Russell indices that it is closer to a few weeks, and it is probably on the low side if we only focused on MSCI and Russell indices.

It is relatively easy to see how the results scale with most of these parameters. Furthermore, the intent of this work is to capture the order of magnitude of things. We assume that there is effectively only one asset manager (who needs to buy the full reconstitution volume). Any security demand not met by the trader, is assumed to be met at reconstitution by other passive market participants who happen to hold some of the specific security (say multiple such investors that had bought some of this security at some point, and irrespective of the reconstitution event). 

Furthermore, for several of the quantities of interest, e.g., the profit the trader makes or the cost difference between a certain asset manager's early trading strategy compared to $Cost_{bench}$ (in US) are independent of, e.g., $S_0$. The same is the case for $\epsilon$ for the asset manager - but not for the trader since the trader first buys and then sells, thus paying the bid-ask spread of $2 \times \epsilon$ (see Eq. ~\ref{eq3}). However, $S_0$ matters to the asset manager when we express the USD cost differential as $bps$ versus the benchmark cost. 

In general, we choose parameter values that would discourage early trading. That is why we use price impact parameters that are on the low end, index weight, volatility, and bid-ask spread on the high end. Even though we also show results for different parameter values, our core model uses the values shown in the last column of Table ~\ref{tab:params}.
\section{Results}
\begin{figure}
    \centering
    \includegraphics[width=0.5\textwidth, height=8cm]{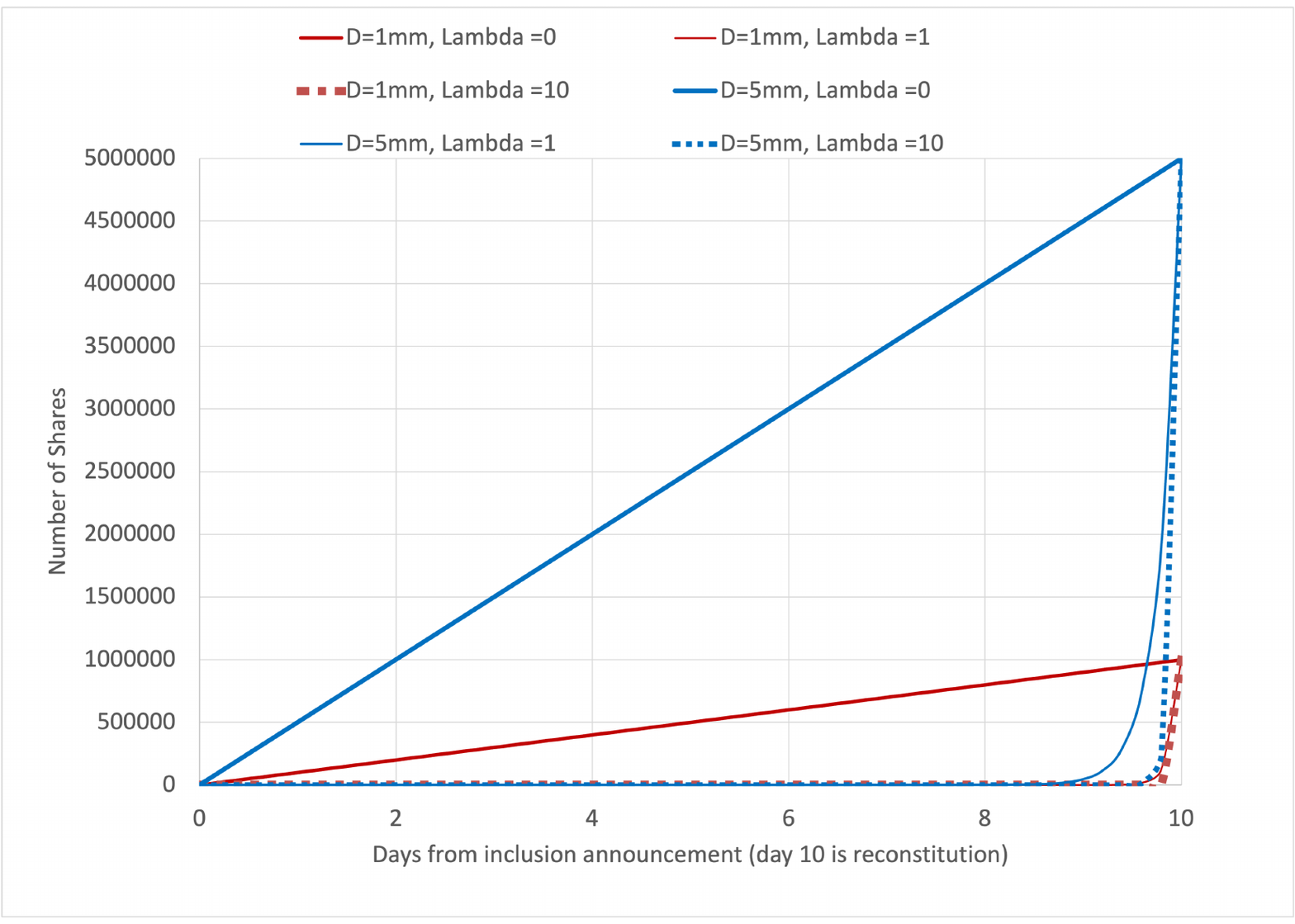}
    \caption{Optimal strategies for share acquisition for the asset manager - no trader assumed - for various values of the tracking error-to-cost parameter ($\lambda$), as well as different levels of total shares needed to meet the index at reconstitution. When concerned only with acquisition cost ($\lambda$ =0), the asset manager adopts a linear rate of buying from announcement to reconstitution ($t=0$ to $t=10$). As emphasis is shifted more and more to minimizing the tracking error (increasing $\lambda$) the "optimal" strategy is to basically wait and buy all shares needed at reconstitution, incurring both higher cost, and possibly, liquidity risk.}
    \label{fig:no_trader}
\end{figure}

\begin{table*}[t]
\centering
\caption{No trader - Impact of $\lambda$ on savings and tracking error under two trade sizes. $\eta = 10^{-6}$, $\gamma = 10^{-8}$, $S_0 = 50$ USD, $\sigma = 30\%$/year, $t_N = 10$ days, $\varepsilon = 0.01$, $w_{\text{bench}} = 0.01$}
\label{tab:notrader}
\begin{tabular}{@{}c ccc ccc@{}}
\toprule
\textbf{$\lambda$} 
& \multicolumn{2}{c}{\textbf{D = 1mm shares}} 
& \phantom{} & \multicolumn{2}{c}{\textbf{D = 5mm shares}} \\
\cmidrule{2-3} \cmidrule{5-6}
& \textbf{Savings (USD)} & \textbf{Tracking Error} 
&& \textbf{Savings (USD)} & \textbf{Tracking Error} \\
\midrule
0   & $\sim$ 1m  ($\sim$ 113bps)    & $\sim$4bps  &&  $\sim$ 14m ($\sim$ 536bps)    & $\sim$4bps \\
0.4 &  $\sim$0m ($\sim$ 0bps)  & $\sim$ 1bps   &&  $\sim$ 1m ($\sim$ 30bps) & $\sim$ 1bps \\ 
10 &  $\sim$0m ($\sim$ 0bps)  & $\sim$ 1bps   &&  $\sim$ 0m ($\sim$ 0bps) & $\sim$ 1bps \\ 

\bottomrule
\end{tabular}
\end{table*}

\subsection{No trader}
As one limit of our toy model, let us consider the case where there is no "predatory trading". In this case, the asset manager still cares about cost and tracking error, but there are no flows $x(t)$ - not beyond ADV in normal markets - that may "force" their hand. 

In this case, we obtain the Euler-Lagrange equation for $\dot{x}, \ddot{x} =0$:
\begin{equation*}
\ddot{y}-\frac{2 \lambda}{\eta D^2} y =0
\end{equation*}

The solution to this differential equation with boundary conditions is as follows:
\begin{itemize}
    \item $y(0) = 0$
    \item $y(t_N) = D$
\end{itemize}
is:
\begin{equation}
y(t) = D \cdot \frac{e^{k t} - e^{-k t}}{e^{k t_N} - e^{-k t_N}}
\end{equation}
where
$$k= \sqrt{\frac{2\lambda}{\eta D^2}}$$
This gives a smooth trajectory that starts at 0 and grows to $D$ over the interval $[0, t_N]$, shaped by the exponential curvature governed by $\lambda$, $\eta$ and $D$.

Figure ~\ref{fig:no_trader} shows the optimal trajectory $y(t)$ for the asset manager for different values of $\lambda$ and $D$ and for $t_N = 10$ days, $S_0 = 50$USD, $\eta = 10^{-6}$, and $\gamma=10^{-7}$. Optimality here mostly refers to with respect to the balancing between cost and tracking error. For $\lambda = 0$, we obtain a straight line - the optimal strategy is to spread trades evenly over time. As we increase $\lambda$, the strategy defers trading to the end of the interval to reduce the tracking error early on. The curve becomes steeper and more loaded as $\lambda$ increases and for the value of 10 almost all trading is delayed until near $t = t_N$, emphasizing the minimal deviation from the index.

The total cost of buying $D$ shares in this case can be shown that it is given by the closed form:
\begin{align*}
\text{Cost} =\ & C_1 \cdot \Bigg[ \frac{(S_0 + \epsilon)}{k} \sinh(k t_N) \\
&\quad + \frac{\gamma D}{\sinh(k t_N)} \cdot \frac{\cosh(2k t_N) - 1}{4k} \\
&\quad + \frac{\eta D k}{\sinh(k t_N)} \left( \frac{t_N}{2} + \frac{\sinh(2k t_N)}{4k} \right) \Bigg]
\end{align*}

with
\[
C_1 = \frac{D \cdot k}{\sinh(k t_N)}
\]

Table ~\ref{tab:notrader} shows savings and tracking errors for some $\lambda$ and $D$ assumptions.  We  see that as we increase $\lambda$ — weighting tracking error more heavily — the total cost increases, converging to the benchmark cost, while tracking error decreases.

\subsection{Linear inventories}
Assume that the trader follows a linear strategy $x(t) = \frac{T}{t_N} t$ and that the manager is willing to trade a fraction $f$ of the total demand $D$ uniformly starting at day d=1, or at day 2, ... N-1, and buys the remaining $(1-f)D$ shares at reconstitution, $t_N$. In other words,
$
y(t) = fDt/d*H(t-d)
$
for $t<t_N$, where $d$ is the day the manager starts buying shares (with values in $[1,N-1]$) and $H$ is the Heaviside function. Using these assumptions, we can derive analytical expressions about the execution cost of the manager for the early execution piece of the trade:
\begin{align*}
\text{Cost} &= \int_d^{t_N} \bigg[ S_0 + \gamma \left( x(t) + y(t) \right) \\
&\quad + \eta \left(\dot{x}(t) + \dot{y}(t)\right) + \epsilon \bigg] \cdot \dot{y}(t) \, dt
\end{align*}
with
\begin{itemize}
    \item $x(t) = \frac{T}{t_N} t$
    \item $\dot{x}(t) = \frac{T}{t_N}$ (constant)
    \item $y(t) = \frac{fD}{t_N - d}(t - d)$ for $t \in [d, t_N]$
    \item $\dot{y}(t) = \frac{fD}{t_N - d}$ over $[d, t_N]$

\end{itemize}
The Heaviside function is responsible for changing the lower limit of integration from $0$ to $d$.

\begin{figure}
    \centering
    \includegraphics[width=0.5\textwidth, height=12cm]{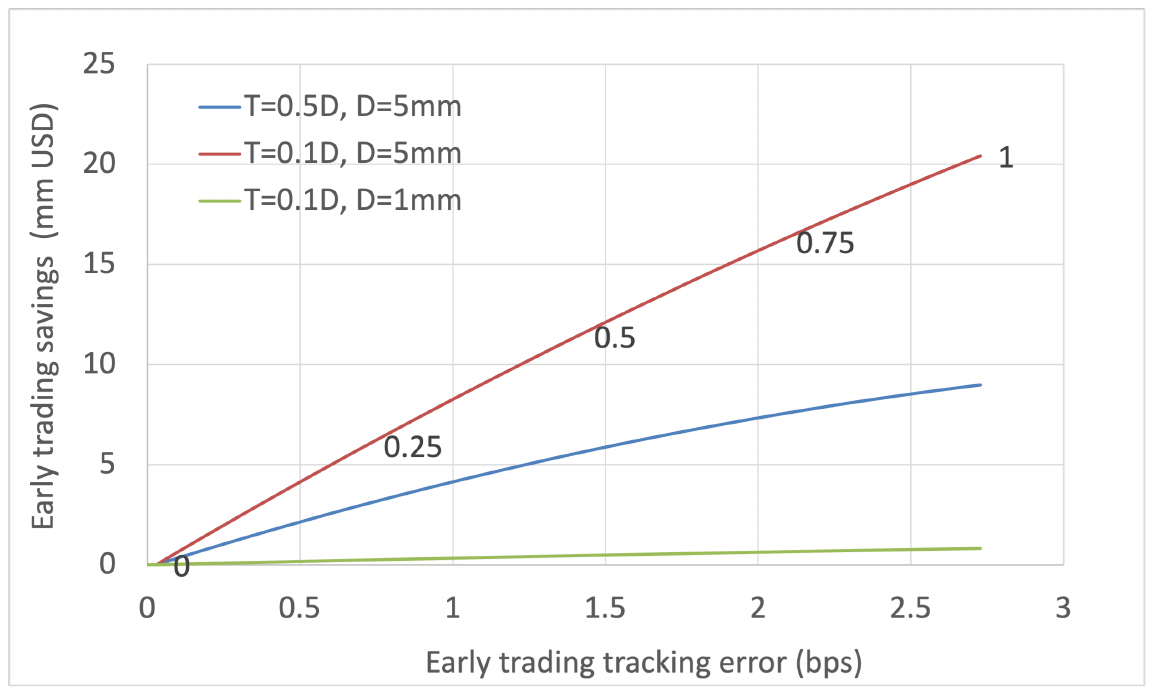}
    \vspace{-4cm}
    \caption{ Early trading savings versus tracking error for the asset manager assuming linear build of inventories for both trader and asset manager. $D$ is the number of share the asset manager will need at reconstitution to meet the benchmark and $T$ is the number of total shares the trader buys and then sells to the asset manager at reconstitution. $t_N=10$, $\eta = 10^{-6}$, $\gamma =10^{-7}$ and $d=1$. Along each curve we vary $f$ - the fraction of $D$ the asset manager will trade early, starting on day $d$ after announcement. Here we assume $d=1$ (yields maximum savings).}
    \label{fig:linear1}
\end{figure}
\begin{figure}
    \centering
    \includegraphics[width=0.5\textwidth, height=12cm]{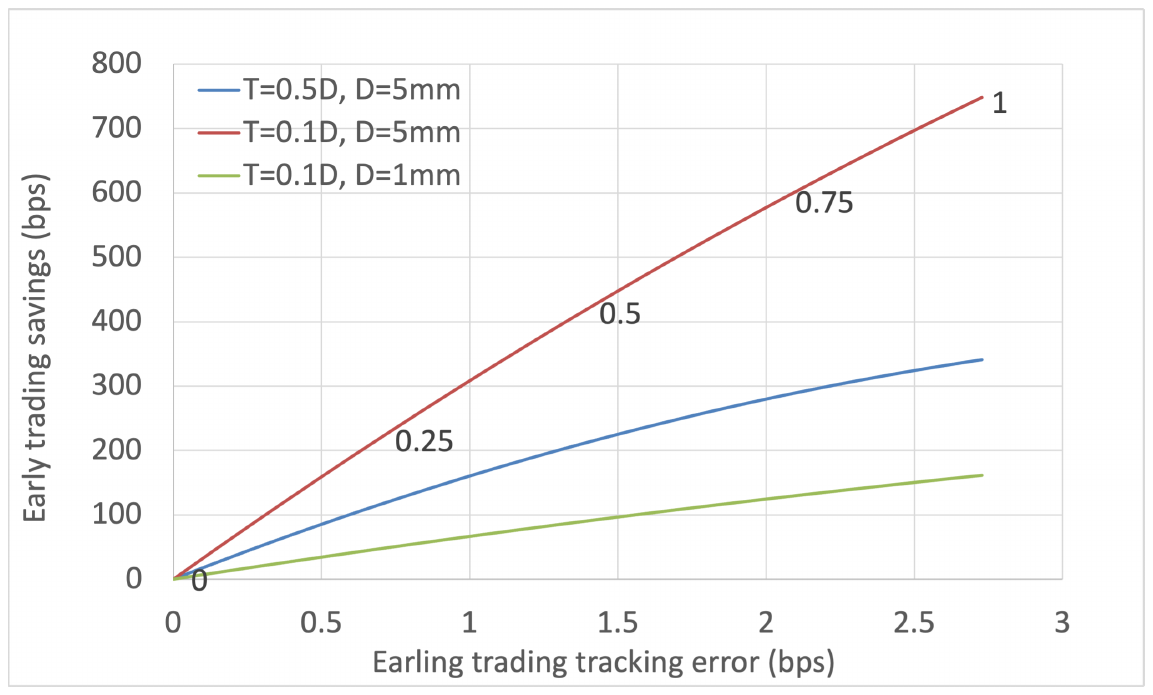}
    \vspace{-4cm}
    \caption{Same as Figure ~\ref{fig:linear1} but with savings in basis points. Note that the mm to basis points transformation is not identical across the 3 sets of parameters, thus the separate Figure (as opposed to adding a second vertical axis in Figure ~\ref{fig:linear1}) }
    \label{fig:savings-bps}
\end{figure}

Substituting the above expressions for $x(t), y(t)$, etc., leads to
\begin{align*}
\text{Cost} = & f D \left[ S_0 + \eta\left( \frac{T}{t_N} + \frac{f D}{t_N - d} \right) + \varepsilon \right] + \\
& 
+ \gamma f D \left[ \frac{T}{2 t_N}(t_N + d) + \frac{f D}{2} \right]
\end{align*}

The overall cost of the strategy is this plus the cost of buying at $t_N$ the remaining $(1-f) D$ shares. This second term is simply equal to $(1-f)$ times the benchmark cost:
$$
Cost_{tot} = Cost + (1-f) Cost_{bench}
$$
with $Cost_{bench}$, as before, the theoretical cost of buying all $D$ shares at the time of reconstitution.

We are interested in estimating (cost) savings because of early execution. This will be equal to the "benchmark" cost of buying all $D$ shares at the time of reconstitution minus the total cost from above:
$$
S_{f} = Cost_{bench} - Cost -(1-f) Cost_{bench}
$$
Substituting in the equation for $S_f$ and doing some calculations leads to our final $S_f$ formula for the early execution cost savings:
\begin{align*}
S_f = & fD \left[
\gamma D + \eta(D - T) - \eta \left( \frac{T}{t_N} + \frac{fD}{t_N - d} \right)
\right]
- \\
& - f \cdot \gamma D \left[ \frac{T}{2} \left(1 + \frac{d}{t_N} \right) + \frac{fD}{2} \right]
\end{align*}

This gives us the cost savings from pre-reconstitution trading. We calculate savings for various assumptions in both $USD$ but also in bps after we normalize the savings $S_f$ by $Cost_{bench}$.

\subsubsection{Tracking Error}

It is straightforward to show that the TEP formula becomes:
\begin{equation}
\Delta TE = w_{bench}(t) \cdot \sigma_{r} \cdot f \cdot \sqrt{\frac{t_N - d}{3}} 
\end{equation}

\subsubsection{Results}
In what follows, we show results for $\epsilon$ = 0.01, $t_N$ = 10 days, $S_0$ = 50 USD. In Figure \ref{fig:linear1} we show savings against TE for $\gamma = 10^{-7}$, $\eta = 10^{-6}$ and a few different assumptions about $T$ and as $D$. We set the day $d$ after announcement that the asset manager starts trading equal to 1 - other values would lead to smaller savings. The value of $f $ranges from 0 to 1 with increasing TE and savings. 
We see that the more shares the trader buys, the higher the market impact and, as a result, the higher the potential for the asset manager to realize savings if they buy $f*D$ shares before reconstitution.

We show similar results in Fig \ref{fig:savings-bps} but for savings in bps (dividing savings by $Cost_{bench}$). We present these results in a different figure, since the mapping between millions and bps is not exactly the same for all parameters combinations. 

\subsection{Games and Optimal Trading Strategies} 
We can consider two game-theoretic frameworks:
\begin{itemize}
\item Simultaneous optimization (Nash): Both players choose paths concurrently.
\item Stackelberg game: the trader leads and the asset manager reacts optimally.
\end{itemize}

\begin{figure}
    \centering
    \includegraphics[width=0.5\textwidth, height=8cm]{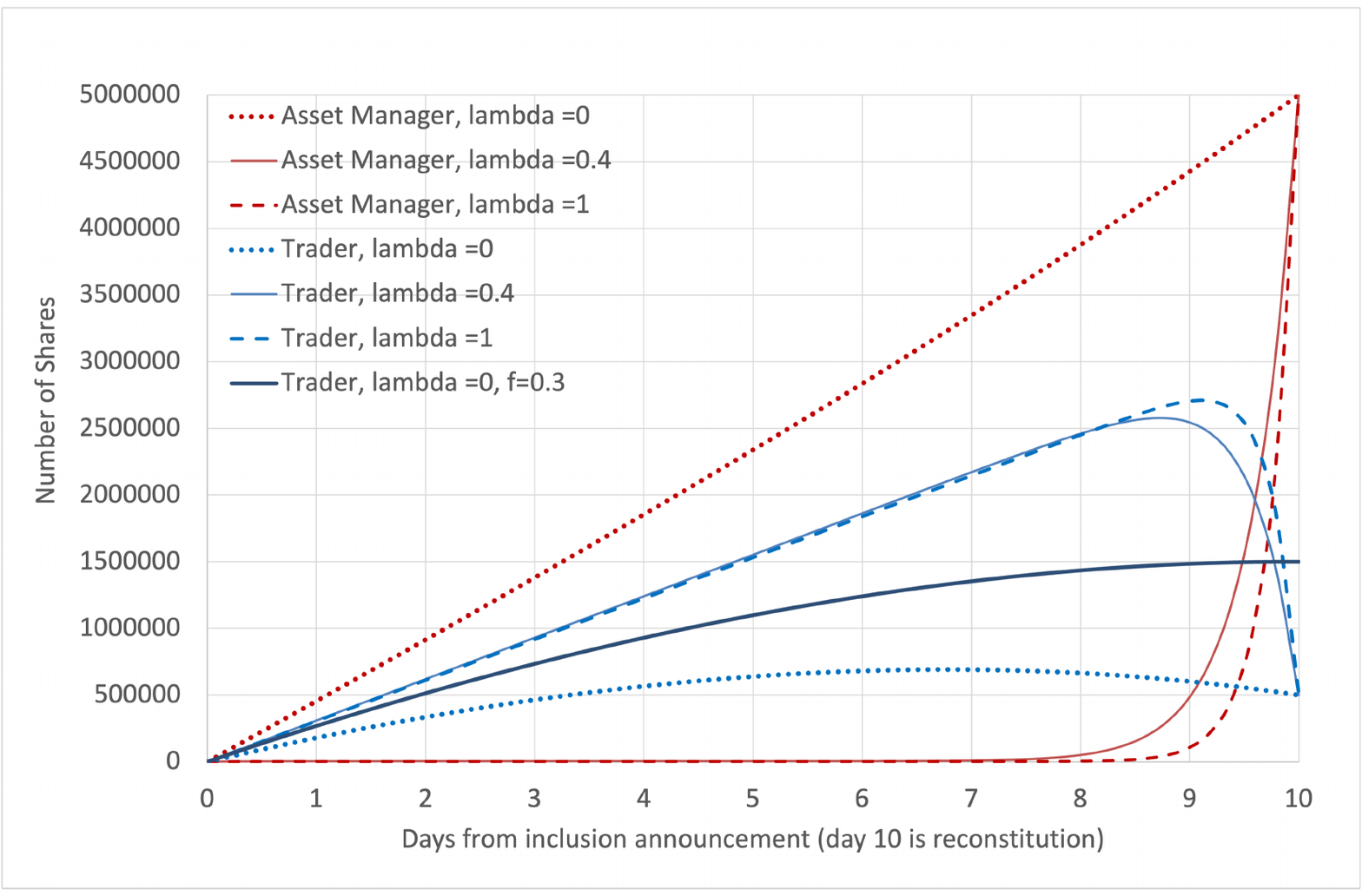}
    \caption{Inventory build trajectories assuming Nash  equilibrium for D=5mm and different $\lambda$ and trader participation rates (f).}
    \label{fig:Nash1}
\end{figure}

\begin{table*}[t]
\centering
\caption{Asset manager cost savings and tracking error, and trader profit for various manager cost-versus-tracking error importance weight ($\lambda$) and total shares needed by the manager ($D$), and for trader participation rates of $f=0.1$ and $f=0.2$ in a Nash game. Manager savings are calculated compared to buying the $D-f*D$ shares throughout the day of reconstitution at a constant rate (of $D-f*D$ per day). Thus, negative savings - when $\lambda$ increases - simply mean that when the asset manager weighs the tracking error more and more, their index change implementation cost will exceed the cost they would incur if instead decided to try and trade throughout the reconstitution day. That is because the more the manager cares about tracking error, the more the trade is pushed as a spike towards the close of the day/moment of index change implementation by the index vendor. }
\label{tab:lambda_dual_trader}
\begin{tabular}{@{}cc cc c cc@{}}
\toprule
\textbf{$\lambda$} & \textbf{D (shares)} 
& \multicolumn{2}{c}{\textbf{Asset Manager}} 
& \phantom{} 
& \multicolumn{2}{c}{\textbf{Trader Profit}} \\
\cmidrule{3-4} \cmidrule{6-7}
&  & \textbf{Savings} & \textbf{Tracking Error} 
&& $f = 0.1$ & $f = 0.2$ \\
\midrule
0   & 5mm & $\sim$21mm ($\sim$754bps)   & $\sim$3bps  && $\sim$2mm ($\sim$788bps)   & --- \\
0.4 & 5mm & $\sim$7mm ($\sim$242bps)   & $\sim$1bps  && $\sim$10mm ($\sim$3571bps) & --- \\
1   & 5mm & $\sim$-2mm ($\sim$-78bps) & $\sim$1bps  && $\sim$14mm ($\sim$5179bps) & --- \\
0   & 1mm & $\sim$1mm ($\sim$162bps)    & $\sim$3bps  && $\sim$0.1mm ($\sim$168bps) & --- \\
0.4 & 1mm & $\sim$-0.5mm ($\sim$-108bps)  & $<$1bps  && $\sim$1mm ($\sim$1454bps)  & $\sim$1mm ($\sim$790bps) \\
1   & 1mm & $\sim$-1mm ($\sim$-257bps)& $<$1bps  && ---                       & $\sim$1mm ($\sim$1163bps) \\
\bottomrule
\end{tabular}
\end{table*}

\subsubsection{Nash Game}
To derive the final equation for the asset manager, let us first solve the trader EL with respect to $\ddot{x}(t)$:
\[
\ddot{x}(t) = -\frac{\eta \ddot{y}(t) + \gamma \dot{y}(t)}{2\eta}
\]

Integrating we obtain:
\begin{equation*}
    \dot{x}(t) = -\frac{1}{2} \dot{y} (t) - \frac{\gamma}{2 \eta} y(t) + K_{1}
\end{equation*}
with $K_{1}$ the integration constant. 

We then substitute this in the manager EL equation, rewriting here for convenience:
\begin{equation*}
\gamma \dot{x}(t) + \eta \ddot{x}(t) + 2\eta \ddot{y}(t) - \frac{2\lambda}{D^2} y (t) = 0
\end{equation*}
and we obtain:
\begin{equation}
\frac{3}{2} \eta \, \ddot{y}(t)
- \gamma \, \dot{y}(t)
- \left( \frac{\gamma^2}{2\eta} 
+ \frac{2\lambda}{D^2}\right) \, y(t)
+ \gamma K_1 = 0
\end{equation}

This is a second-order linear ODE in $y(t)$, with general solution

\begin{equation}
y(t) = y_{p}  
+ C_1 e^{r_1 t} + C_2 e^{r_2 t}
\end{equation}

where:

\begin{equation*}
r_{1,2} = \frac{\gamma}{3\eta} \pm \frac{\sqrt{4\gamma^2 + \frac{12\eta\lambda}{D^2}}}{3\eta}
\end{equation*}

and 

\begin{equation*}
y_p(t) = \frac{\gamma K_1}{\frac{\gamma^2}{2 \eta} + \frac{2 \lambda}{D^2}}
\end{equation*}
the particular solution we obtain if we assume that $y_p$ is a constant, 
and $C_1, C_2$ are constants that we set through the boundary conditions:

\begin{align*}
C_1 &= \frac{D + y_p (C - 1)}{B - C} \\
C_2 &= \frac{-D - y_p (B - 1)}{B - C}
\end{align*}
with
\begin{align*}
B &= e^{r_1 t_N} \\
C &= e^{r_2 t_N}
\end{align*}

Above and in what follows we impose the boundary conditions:
\begin{align*}
y(0) &= 0 \\
y(t_N) &= D \\
x(0) & = 0 \\
x(t_N) & = T
\end{align*}
Note that we determine $K_1$ from the trader boundary conditions and then substitute in $y_p$.

\subsubsection{Trader}
The solution for $x(t)$, given the equation:

$$
\dot{x}(t) = -\frac{1}{2} \dot{y}(t) - \frac{\gamma}{2\eta} y(t) + K_1
$$

and using:

$$
y(t) = y_p + C_1 e^{r_1 t} + C_2 e^{r_2 t}
$$

is:
\begin{align*}
x(t) & = -\left( \frac{1}{2} + \frac{\gamma}{2\eta r_1} \right) C_1 \left( e^{r_1 t} - 1 \right) \\
    &    -\left( \frac{1}{2} + \frac{\gamma}{2\eta r_2} \right) C_2 \left( e^{r_2 t} - 1 \right) \\
    &    - \frac{\gamma}{2\eta} y_p t + K_1 t
\end{align*}
We set the value of $K_1$ by the condition $x(t_N)=T$.

We show results on the inventory paths in Figure ~\ref{fig:Nash1}. Interestingly, when the asset manager is more eager to wait - weighing more heavily the small tracking error than the smaller cost - the trader seems to try to manipulate the circumstances by buying more than $T$ to push the price as high as she can, before starting to sell a day or 2 before reconstitution. We allowed the solution to exhibit such behavior, even though that assumes that at some point the trader may need to have funds to buy many more than $T$ shares.

Apparently, this manipulation can bring the cost for the asset manager to even higher than the "all at once" benchmark cost. That is why in Table ~\ref{tab:lambda_dual_trader} we see for higher $\lambda$ values slightly negative "savings" for the asset manager (i.e., she ends up paying a bit more than the case where there was not market activity at all and then at reconstitution she appeared trying in buy $D$ shares with $T$ shares appearing). 
\begin{figure}
    \centering
    \includegraphics[width=0.5\textwidth, height=8cm]{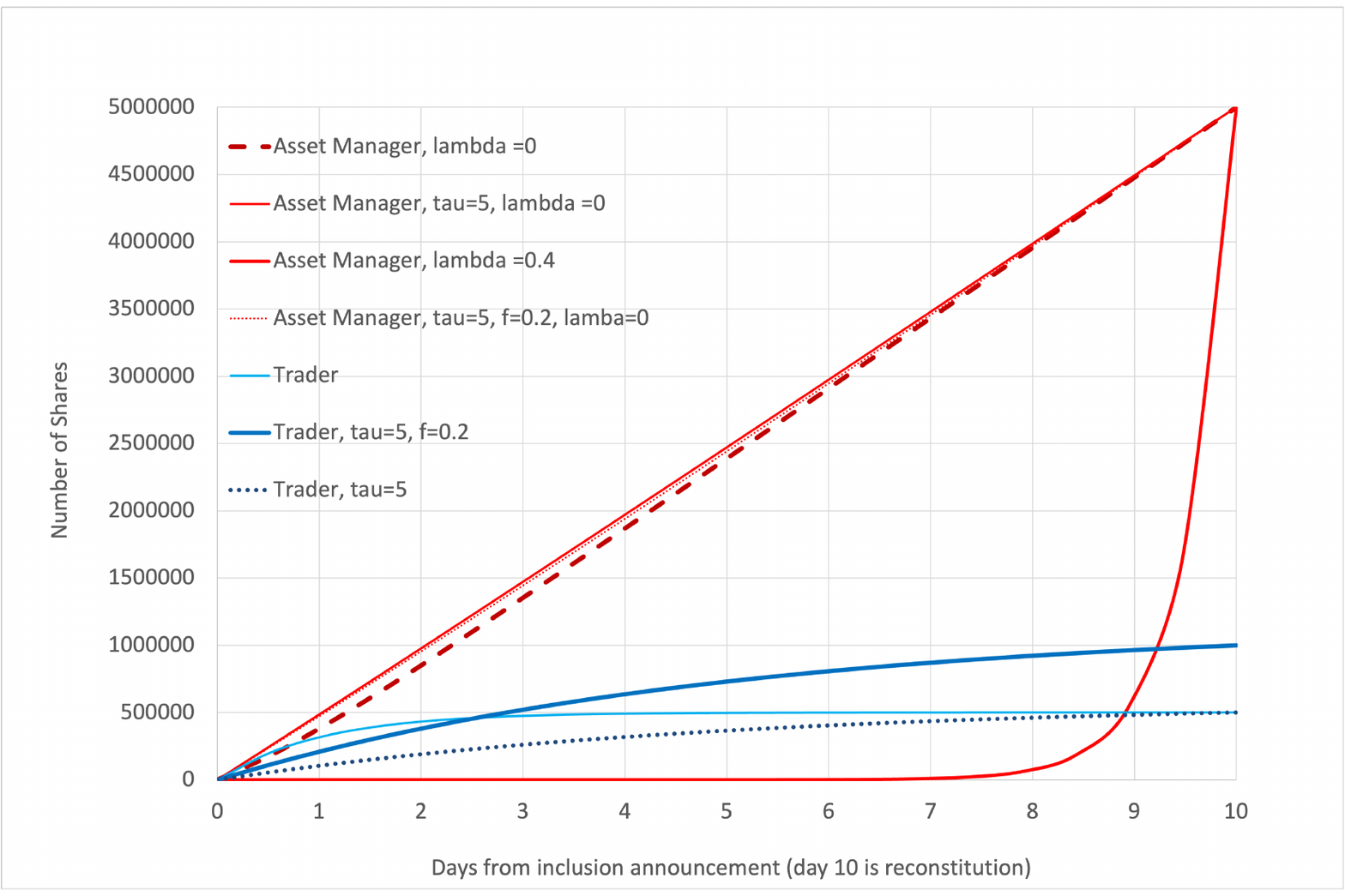}
    \caption{Inventory build trajectories for the asset manager and the trader assuming a Stackelberg game. Note that the trader trajectory does not depend on $\lambda$ in this case. Unless otherwise indicated, the results assume $f=0.1, \tau=1$.}
    \label{fig:stack}
\end{figure}
\begin{table*}[t]
\centering
\caption{Asset manager potential cost savings, and tracking error and trader profit for different values of $\lambda, D, f$ and $\tau$ assuming a Stackelberg game in which the trader picks her strategy and the asset manager optimizes in response to the trader choice. Manager savings are calculated compared to buying the $D-f*D$ shares throughout the day of reconstitution at a constant rate (of $D-f*D$ per day). Thus, negative savings - when $\lambda$ increases - simply mean that when the asset manager weighs the tracking error more and more, their index change implementation cost will exceed the cost they would incur if instead decided to try and trade throughout the reconstitution day. That is because the more the manager cares about tracking error, the more the trade is pushed as a spike towards the close of the day/moment of index change implementation by the index vendor.}
\label{t4}
\begin{tabular}{@{}c@{\hskip 10pt}c@{\hskip 15pt}c@{\hskip 10pt}c@{\hskip 10pt}c@{\hskip 10pt}c@{\hskip 10pt}c@{}}
\toprule
$\lambda$ & D (shares) & \textbf{Savings} & \textbf{Tracking Error} & $f$ & $\tau$ & \textbf{Profit} \\
\midrule
0   & 5mm  & $\sim$21mm ($\sim$ 756bps)     & $\sim$3bps & 0.1 & 1 & $\sim$2mm ($\sim$ 840bps) \\
0.4 & 5mm  & $\sim$-2mm ($\sim$ -100bps)  & $\sim$1bps & 0.1 & 1 & $\sim$2 mm ($\sim$ 945bps) \\
0   & 5mm  & $\sim$18mm ($\sim$659bps)     & $\sim$3bps & 0.2 & 5 & $\sim$3.5mm ($\sim$715bps) \\
0   & 1mm  & $\sim$1mm ($\sim$ 163bps)    & $\sim$3bps & 0.1 & 1 & $\sim$0.1mm ($\sim$166bps) \\
0.4 & 1mm  & $\sim$-1mm ($\sim$ -258bps)   & $\sim$1bps & 0.1 & 1 & $\sim$0.1mm ($\sim$185bps) \\
0   & 1mm  & $\sim$1mm ($\sim$ 140bps)    & $\sim$3bps & 0.2 & 1 & $\sim$0.1mm ($\sim$140bps) \\
0.4 & 1mm  & $\sim$-1mm ($\sim$ -280bps)   & $\sim$1bps & 0.2 & 1 & $\sim$0.1mm ($\sim$154bps) \\
0   & 1mm  & $\sim$1mm ($\sim$ 141bps)    & $\sim$3bps & 0.2 & 5 & $\sim$0.1mm ($\sim$142bps) \\
\bottomrule
\end{tabular}
\end{table*}

\subsubsection{Stackelberg game}
In this setup, the trader ("leader") chooses $x(t)$ first and the manager ("follower") chooses $y(t)$ to minimize her own cost, given $x(t)$. 

Let us assume that the trader wants to buy fast after announcement to leverage the lower prices and then decelerate. We assume that the purchase plan is given by:
\begin{equation}
x(t) = \frac{T}{(1-e^{-t_N/\tau})} (1-e^{-t/\tau})
\end{equation}
and, as before, $T$ is the total number of shares the trader has acquired by $t_N$, the reconstitution day/time. The parameter $\tau$ determines how front-loaded the inventory building path is: smaller $\tau$ values result in a steep initial rise, plateauing early. Larger values spread the growth more evenly, with a slower climb.

Substituting into the asset manager EL equation:
\[
\frac{C}{\tau} \left( \gamma - \frac{\eta}{\tau} \right) e^{-t/\tau}
+ 2\eta \ddot{y}(t)
- \frac{2\lambda}{D^2} y(t)
= 0
\]
with $C= \frac{T}{(1-e^{-t_N/\tau})}$. Let \( k = \sqrt{\frac{\lambda}{\eta D^2}} \). Then the solution is:
\begin{equation}
y(t) = C_1 e^{kt} + C_2 e^{-kt} + y_p
\end{equation}
with
\begin{equation*}
y_p = K e^{-t/\tau}
\end{equation*}
\[
K = \frac{ \frac{\eta C}{\tau^2} - \frac{\gamma C}{\tau} }{ \frac{2\eta}{\tau^2} - \frac{2\lambda}{D^2}}
\]

and the integration constants \( C_1 \) and \( C_2 \) are given by:
\[
C_1 = \frac{D - K(E - B)}{A - B}, \quad
C_2 = \frac{-D - K(A - E)}{A - B}
\]

\[\quad A = e^{k t_N}, \quad B = e^{-k t_N}, \quad E = e^{-t_N/\tau}
\]

We show results for the asset manager savings and tracking error, as well as the trader profits for various parameter assumptions in Table ~\ref{t4}, whereas inventory path examples are shown in Fig. ~\ref{fig:stack}.

\section{Discussion and Conclusion}
\begin{figure}
    \centering
    \includegraphics[width=0.5\textwidth, height=6cm]{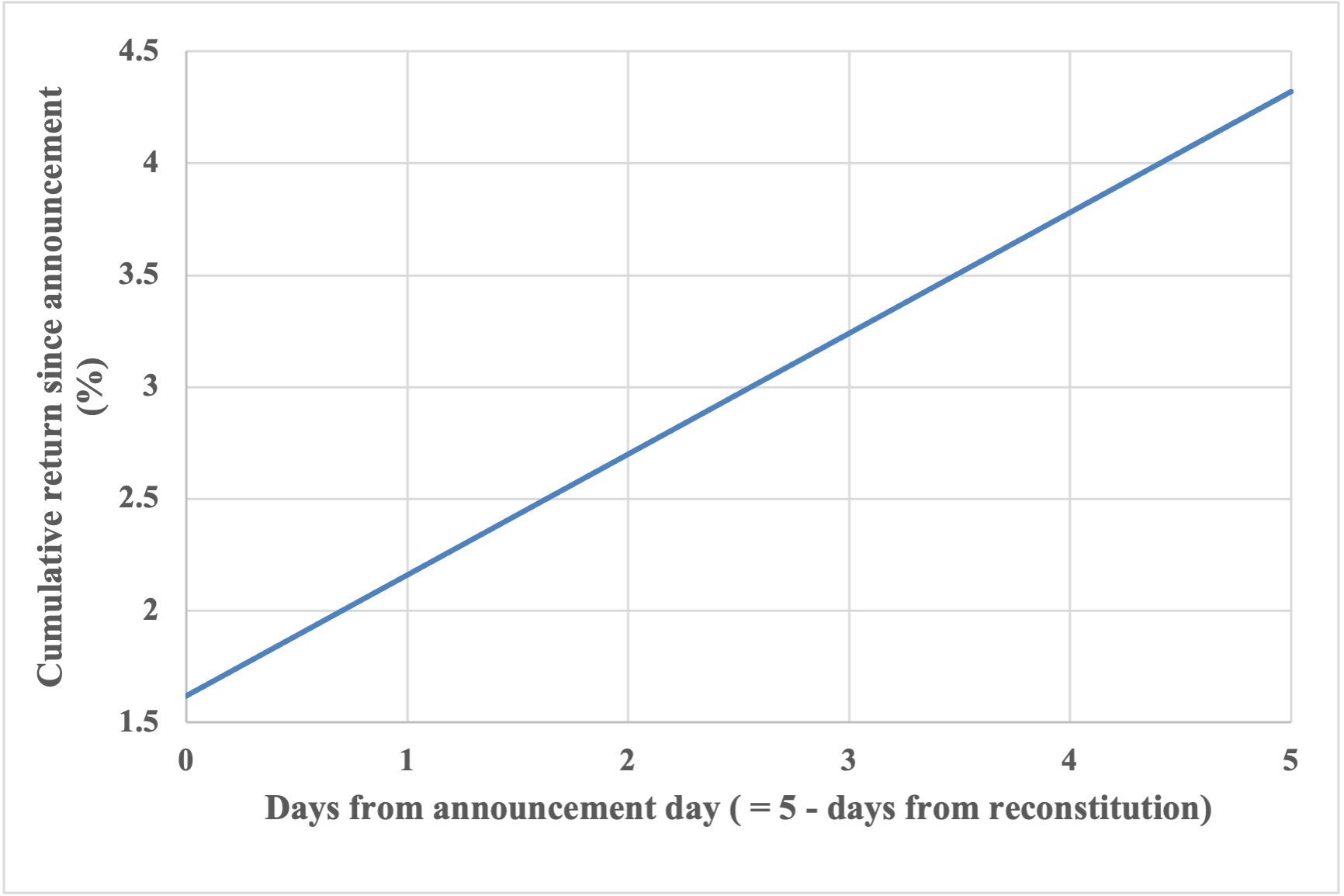}
    \caption{Real world data show that prices of stocks rise between their index inclusion announcement and the inclusion date implementation. The magnitudes of these observed effects are easily reproduced by the models discussed in this work.}
    \label{fig:only_trader}
\end{figure}

Index inclusion can create short-term mispricing opportunities due to inelastic demand from passive funds. Traders can profitably anticipate index flows by building positions early and providing liquidity near the close on reconstitution day. This can exert pressure on the stock prices as has been shown in various studies based on real data. The models and parameters discussed here can offer plausible estimates of such effects. For example, assuming we just have one trader buying before reconstitution, the path the price will follow (assuming the trader optimizes for her profit) for $T=0.9\cdot D, D=\rm{5mm}, t_{N}=5 \rm{days}, \gamma = 3 \times 10^{-7}$ and $\eta = 10^{-6}$ is shown in Fig.~\ref{fig:only_trader}. In this case we tried to emulate without much experimentation the $\sim 5\%$ cumulative result between announcement and reconstitution reported in the observational study of \cite{arnott}.

For passive managers, delayed execution of reconstitution trades —especially at the closing auction of the reconstitution day—can result in substantial increase of the implementation costs. Executing trades gradually or in advance (even fractionally) may reduce these costs without materially affecting tracking error which is often the reason for the delayed trading.  But, how sensitive is the tracking error to a few days of deviation from the index?

Let's assume that we are calculating the ex-post tracking error using daily returns and a $252$ business days rolling window. The tracking error then will be:
\[
TE^2 =\frac{1}{252} \sum_{i=1}^{252} \left[ \left(r_{f, i} - r_{\mathcal{I}, i} \right) - \left( \bar{r}_f - \bar{r}_{\mathcal{I}} \right) \right]^2
\]
with the index $'f'$ referring to the passive fund, and $'I'$ referring to the market index the fund follows. 

Suppose that over a period of $t_N$ days there is a $\delta r$ return adjustment/difference between the fund and the index:
\begin{align*}
TE'^2 & = \frac{1}{252}[\sum_{i=1}^{252-t_N} \left[ \left(r_{f, i} - r_{\mathcal{I}, i} \right) - \left( \bar{r}_f - \bar{r}_{\mathcal{I}} \right) \right]^2 + \\
& \sum_{i=252-t_N+1}^{252} \left[ \left(r_{f, i} + \delta r - r_{\mathcal{I}, i} \right) - \left( \bar{r}_f - \bar{r}_{\mathcal{I}} \right)\right]^{2}  ] \\
 & = TE^2 +  \\
 & \frac{1}{252} \sum_{i=252-t_N+1}^{N} \left[ \delta r^2 - 2\delta r \cdot  r_{f,i} + 2\delta r \cdot r_{\mathcal{I},i} \right]
\end{align*}

For simplicity we assume that $t_N$ is the most recent $t_N$ days and that the $\delta r$ has an average of 0 without loss of generality. Assuming that $\delta r$ is uncorrelated to either returns, this simplifies to:
\[
TE^{'2}= TE^2 + \frac{t_N}{252}\cdot \langle \delta r^2 \rangle
\]
In other words, any additional tracking error introduced by early trading would be
\begin{equation}
    \Delta TE = \sqrt{TE^2 + \frac{t_N}{252}\cdot \langle \delta r^2 \rangle}-TE
\end{equation}
Any conclusions will depend on the specifics, but it is clear that the short duration of realizing TE due to early trading, along with the potential impacts on cost implementation we have seen, would make a case that the "right" strategy may not be "buy everything you need early" but it is not "wait and buy everything at the market close the day of the index reconstitution", either.

Similarly, we can estimate the drag exerted on the average annual return of a fund by the complete avoidance of tracking error: in Tables 3 and 4, the \emph{Savings} column is the cost the manager would save by trading early relative to buying everything at reconstitution(the zero-TE benchmark). If the manager focuses on avoidance of tracking error, she forgoes those savings and the drag from zero-TE is 

\begin{equation*}
\text{Drag (bps)} = \frac{\text{Savings (USD)}}{\text{Fund AUM (USD)}} \times 10{,}000.
\end{equation*}

Summing across all events/names in the year gives the annual drag. For example, Table 3 (first row) shows savings of $\simeq 21mm$ for 5mm shares at $\lambda=0$. For a fund AUM of $50B$, the per-event drag from avoiding TE is:
\begin{equation*}
\frac{21mm}{50000mm} \times 10000 \approx 4.2 \text{bps}.
\end{equation*}
If this happens a few times (multiple names and/or reconstitutions) per year this drag can become substantial.

Although not explicitly accounted for in this work, it is important to note that purchasing all shares at the time of reconstitution introduces an additional source of loss beyond the pre-reconstitution effects discussed. Specifically, any difference between the inflated purchase price, $S_{t_N}$, and the post-reconstitution price, $S_a$ due to post-reconstitution price reversal, represents an incremental loss in value. We will include this in a future work.

\section{Appendix}
\subsection{Stackelberg game closed form solutions}
\subsubsection{Manager Cost}
\begin{equation*}
\int_0^{t_N} \left( S_0 + \gamma(x + y) + \eta(\dot{x} + \dot{y}) + \epsilon \right) \dot{y}(t) \, dt = \sum_{i=1}^6 \text{Term}_i
\end{equation*}
Each term is:
\begin{align*}
\text{Term}_1 &= S_0 \left[ C_1 (e^{kt_N} - 1) - C_2 (1 - e^{-kt_N}) - K (1 - e^{-t_N/\tau}) \right] \\
\text{Term}_2 &= \gamma C C_1 k \left[ \frac{1}{k}(e^{kt_N} - 1) - \frac{1}{k - \frac{1}{\tau}} (e^{(k - \frac{1}{\tau}) t_N} - 1) \right] \\
&\quad - \gamma C C_2 k \left[ \frac{1}{k}(1 - e^{-kt_N}) - \frac{1}{k + \frac{1}{\tau}} (1 - e^{-(k + \frac{1}{\tau}) t_N}) \right] \\
&\quad - \gamma C K \left[ (1 - e^{-t_N/\tau}) - \frac{1}{2}(1 - e^{-2t_N/\tau}) \right] \\
\text{Term}_3 &= \frac{\gamma C_1^2}{2}(e^{2kt_N} - 1)  \\
& -\frac{\gamma C_2^2}{2}(1 - e^{-2kt_N}) - \frac{\gamma K^2}{2}(1 - e^{-2t_N/\tau}) \\
&\quad - \gamma C_1 K \frac{k}{\tau(k - \frac{1}{\tau})}(e^{(k - \frac{1}{\tau})t_N} - 1) \\
&\quad - \gamma C_2 K \frac{k}{\tau(k + \frac{1}{\tau})}(1 - e^{-(k + \frac{1}{\tau})t_N}) \\
\text{Term}_4 &= \eta A C_1 k \frac{1}{k - \frac{1}{\tau}}(e^{(k - \frac{1}{\tau})t_N} - 1) \\
& -\eta A C_2 k \frac{1}{k + \frac{1}{\tau}}(1 - e^{-(k + \frac{1}{\tau})t_N}) \\
&\quad - \frac{\eta A K}{2}(1 - e^{-2t_N/\tau}) \\
\text{Term}_5 &= \frac{\eta C_1^2 k}{2}(e^{2kt_N} - 1)  \\
& +\frac{\eta C_2^2 k}{2}(1 - e^{-2kt_N}) + \frac{\eta K^2}{2\tau}(1 - e^{-2t_N/\tau}) \\
&\quad - 2\eta C_1 C_2 k^2 t_N - 2\eta C_1 k K \frac{1}{\tau(k - \frac{1}{\tau})}(e^{(k - \frac{1}{\tau})t_N} - 1) \\
&\quad + 2\eta C_2 k K \frac{1}{\tau(k + \frac{1}{\tau})}(1 - e^{-(k + \frac{1}{\tau})t_N}) \\
\text{Term}_6 &= \epsilon \left[ C_1 (e^{kt_N} - 1) - C_2 (1 - e^{-kt_N}) - K (1 - e^{-t_N/\tau}) \right]
\end{align*}
\subsubsection{Trader Cost}
\[
\begin{aligned}
\int_0^t \left( S_0 + \gamma(x + y) + \eta(\dot{x} + \dot{y}) + \epsilon \right) \dot{x}(t) \, dt =\;&
A_0 C_x \tau \left(1 - e^{-t/\tau} \right) \\
&+ A_1 C_x \frac{e^{(k - \frac{1}{\tau})t} - 1}{k - \frac{1}{\tau}} \\
&+ A_2 C_x \frac{1 - e^{-(k + \frac{1}{\tau})t}}{k + \frac{1}{\tau}} \\
&+ A_3 C_x \frac{\tau}{2} \left(1 - e^{-2t/\tau} \right)
\end{aligned}
\]
\hspace{2cm} With:
\[
\begin{aligned}
 & & A_0 = S_0 + \gamma C + \epsilon \\
& &  A_1 = \gamma C_1 + \eta C_1 k \\
& & A_2 = \gamma C_2 - \eta C_2 k \\
& & A_3 = (K - C)\left( \gamma - \frac{\eta}{\tau} \right) \\
& &  C_x = \frac{C}{\tau}\\
& & \quad k = \sqrt{\frac{\lambda}{\eta D^2}}\\
& & C = \frac{T}{1 - e^{-t_N/\tau}}\\
\end{aligned}
\]
\end{document}